\documentclass[reprint,amsmath,amssymb,aps]{revtex4-1}
\usepackage{graphicx}
\usepackage{bm}
\usepackage[utf8]{inputenc}
\usepackage{lineno,hyperref}
\modulolinenumbers[5]
\usepackage{amsfonts}
\numberwithin{equation}{section}
\usepackage{mathtools}

\begin{document}

\title{Spinning closed superstring: as Nature's building block}
\author{Hongsu Kim}
\email{chris@kasi.re.kr}
\affiliation{Center for Theoretical Astronomy, Korea Astronomy and Space Science Institute, Daejeon 34055, Republic of Korea}
\date{\today}

\begin{abstract}
Ever since its birth, up until its present development, the major role of string theory involves being the best candidate for the theory of quantum gravity and other species of interactions. In the present work, we would like to accomplish this goal by minimally extending its content while greatly simplifying its structure. To be more specific, by endowing the closed superstring with its spin, we successfully achieve this goal. This issue has been addressed in the first part of this work, entitled, "mission 1 of the work". In addition, we would like to make further developments on the selection of compactification manifolds that brings the string theory from the critical 10 dimensions down to four cosmology dimensions. Indeed, this issue has not been fully seriously and extensively explored in the literature. In the present work, therefore, we want to bring serious attention to this missing but non-trivial issue. This issue has been addressed in the second part of this work, entitled, "mission 2 of the work".
\begin{description}
\item[PACS numbers]
\item[Keywords]
\end{description} 
\end{abstract}
 
\maketitle

\section{MISSION 1 OF THE WORK}
\subsection{Introduction}
 \subsubsection{[Breakthrough]1}
 As long as the closed superstring is allowed to spin, generically it involves (non-abelian) gauge mode as well as the gravity mode and hence one does not need to introduce the Dirichlet brane (D-brane) in addition to the F1-brane (string) just to involve (non-abelian) gauge field mode. That is, the spin mode that the closed superstring (but not the open superstring) exclusively possesses, completely replaces the quest for D-brane mostly to involve (non-abelian) gauge field/theories in the string phenomenology. In short, the spin of closed superstring mostly replaces the quest for D-brane in the latest development of superstring theory [2nd string revolution]. To conclude, therefore, open superstring perturbation (vibration) can only produce gauge fields. However spinning closed superstring perturbations \{vibration \& spin (winding)\} may well produce both gravity (=geometry) and gauge field. 
    
 \subsubsection{[Breakthrough]2}
In this new paradigm the spin of closed superstring appears to replace the quest for D-brane and take over its role as far as the advent of (non-abelian) gauge theory is concerned. This state of the affair, then, may pose the concern: what about the gauge/gravity duality as a consequence of brane/bulk duality that excited the entire theoretical physics community for the past 20 years or so? Yes, this business is safely preserved by the following rationale: due to the presence of both vibrational and spin modes together, the string tension and the centrifugal force compete and affect each other all the time. As a result, the (non-abelian) gauge theory and the gravity associated with spin and vibrational modes respectively affect each other all the time and hence this coupling replaces the gauge/gravity duality in the context of brane/bulk duality.
That is, regardless of the presence/absence of D-brane (by J. Polchinsky), or equivalently, Regardless of the open/closed string duality, “spinning-closed superstring” alone may well be the fundamental constituent of nature.
In short, spinning-closed superstring alone = \{gravity \& all species of gauge theories\}.

\subsection{Strategy}
\begin{figure}[h!]
\includegraphics[width=8.5cm]{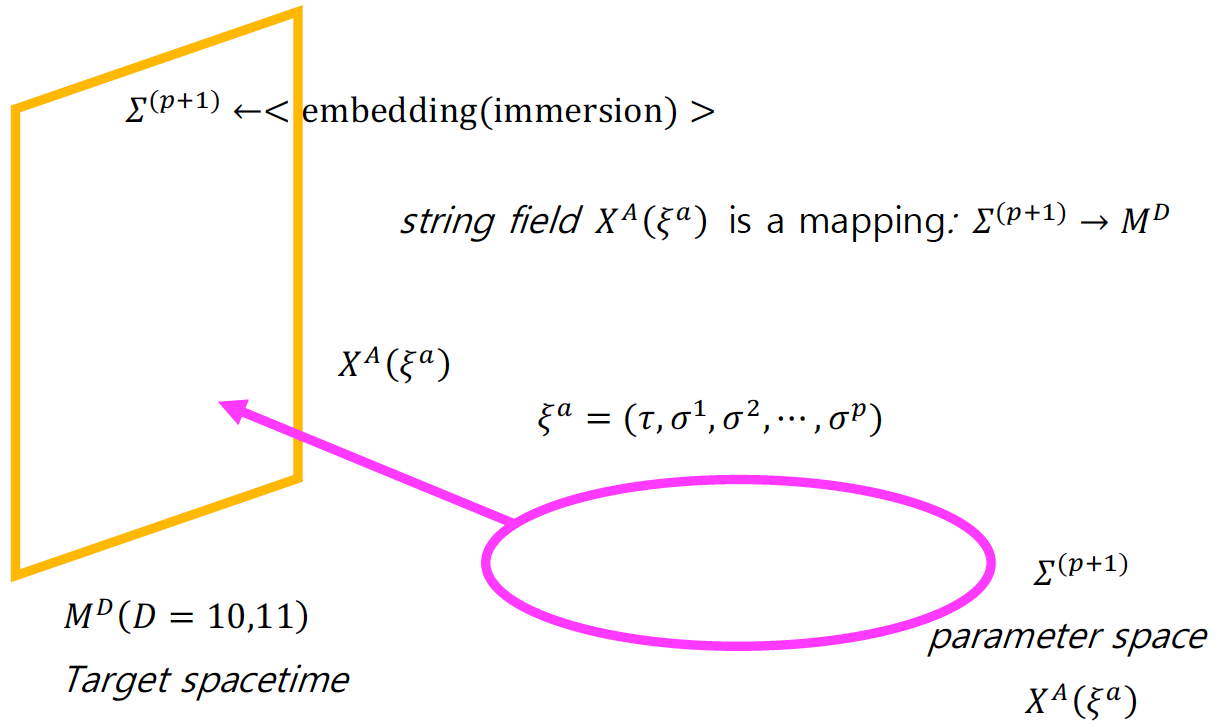}
\caption{string coordinate or string field is a mapping from the parameter space to the string target spacetime. As a result, the image that string field draws naturally becomes a curve with respect to which one can define both its tangent and unit normal vectors. (For the case at hand $p=1$, namely the string)}
\label{fig:1}
\end{figure}

$p$-brane world volume coordinate is an embedding (immersion) of ($p$+1)-dim. parameter space into D-dim. target spacetime. Pullback of target spacetime metric $G_{AB}$ induced metric onto the world volume of $p$-brane
\begin{equation}
g_{ab}=G_{AB}\frac{\partial X^A}{\partial \xi^a}\frac{\partial X^B}{\partial \xi^b}.
\end{equation}
Taking a "static gauge" in which
\begin{align}
\tau&=x^0,\ \sigma^i=x^i (i=1, 2, \cdots, p)\\
g_{\mu\nu}&=G_{AB}\frac{\partial X^A}{\partial x^\mu}\frac{\partial X^B}{\partial x^\nu}
\end{align}
Tangent vectors on $\Sigma^{(p+1)}$: ($p+1$)-unit tangents\\
(static gauge) $t^A_\mu=\frac{\partial X^A}{\partial x^\mu} (a,\mu=0, 1, \cdots, p)$ 

Normal vectors [$D-(p+1)$]-unit normal $n^A_\alpha$ orthogonal to ($p+1$)-tangents at each $X^A(\xi^a)$ on $\Sigma^{(p+1)}\ (a = (p+1),\cdots, D)$ satisfying the orthogonality
\begin{align}
n^A_\alpha n^A_\beta&=\delta_{\alpha\beta}\\
t^A_\mu t^A_\nu&=\frac{\partial X^A}{\partial x^\mu}\frac{\partial X^A}{\partial x^\nu}=g_{\mu\nu}\\
n^A_\alpha n^A_\mu&=0
\end{align}
thus $\{n^a_\alpha, t^A_\mu\}$ form basis D-vectors in target spacetime and the basic equations of embedding are
\begin{align} 
   \nabla_\mu t^A_\nu&=\partial_\mu t^A_\nu-\Gamma^\lambda_{\mu\nu}t^A_\lambda=K^\alpha_{\mu\nu}n^A_\alpha, \\
   D_\mu n^A_\alpha&=\partial_\mu n^A_\alpha+A^\alpha_{\mu\beta}n^A_\beta=-K^\nu_{\alpha\mu}t^A_\nu
  \end{align}
 where $\nabla_\mu$ is geometry covariant derivative of tangent vector to closed superstring in the bulk and
 $D_\mu$ is gauge covariant derivative of normal vector to open superstring on the brane.
$A_{\mu\alpha\beta}$ is a $S^D(D-(p+1))$ connection for the parallel transport of the unit normal vector $n^A_\alpha$ on the brane.

\begin{figure}[h!] 
\includegraphics[width=8cm]{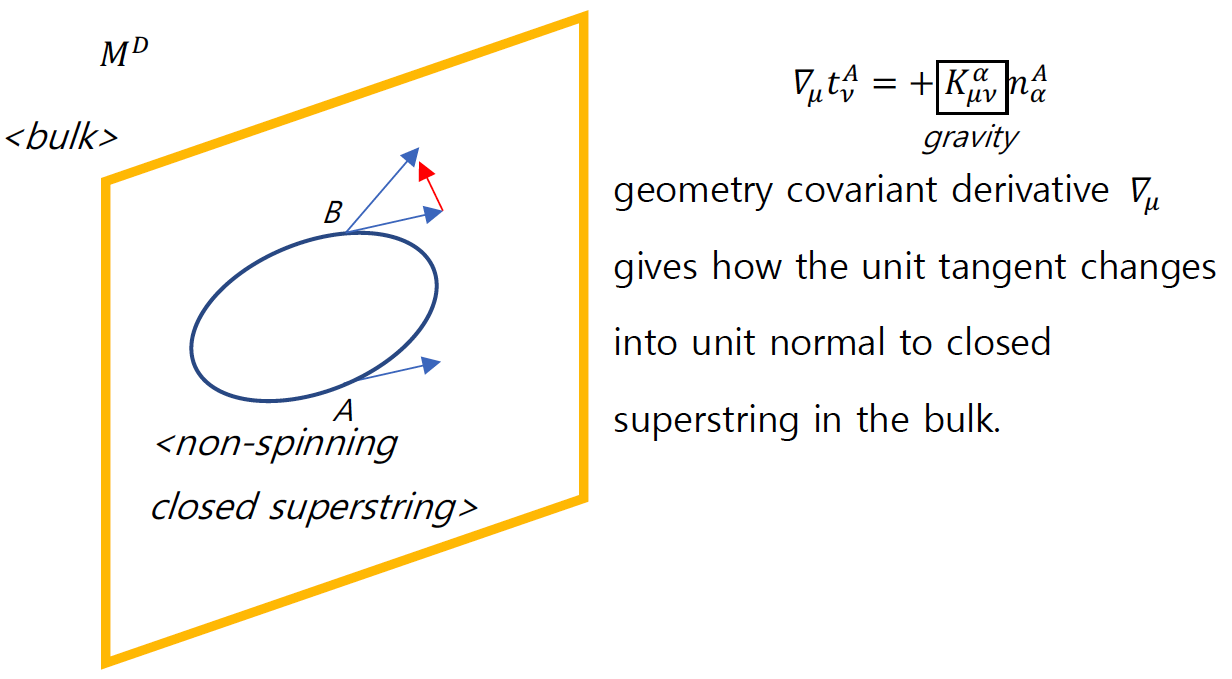}
\caption{perturbations on a non-spinning closed superstring produce gravity alone.
\newline Slightly along the non-spinning closed superstring unit normal remains the same. However, unit tangent may undergo some discrepancy/deviation which can be identified with unit normal times some tensor coefficient. And this tensor coefficient involves affine connection which, in turn, generates spacetime geometry, namely, gravitation.}
\label{fig:2}
\end{figure}

\begin{figure}[h!]
\includegraphics[width=9cm]{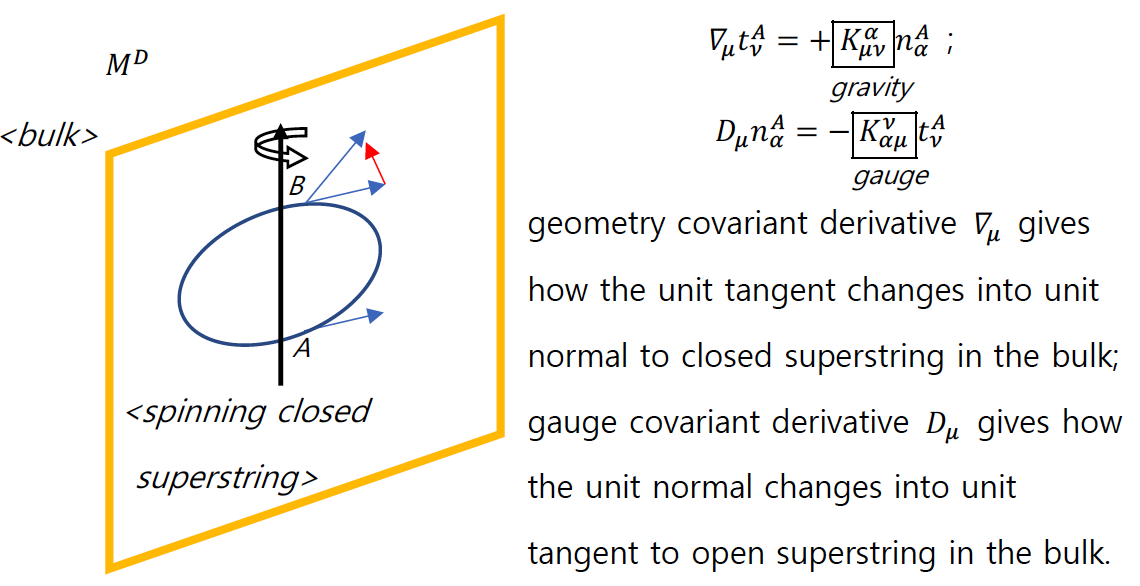}
\caption{perturbations on a spinning closed superstring produce both gravity and gauge theory.
\newline Unlike the case of non-spinning closed superstring, slightly along the spinning closed superstring, unit normal may undergo some discrepancy/deviation which can be identified with unit tangent times some tensor coefficient. And this tensor coefficient involves gauge connection which, in turn, generates (non)abelian gauge theory. Meanwhile, unit tangent undergoes some discrepancy/deviation which can be identified with unit normal times some tensor coefficient. And this tensor coefficient involves affine connection which, in turn, generates spacetime geometry, namely, gravitation.}
\label{fig:3}
\end{figure}

\begin{figure}[h!]
\includegraphics[width=9cm]{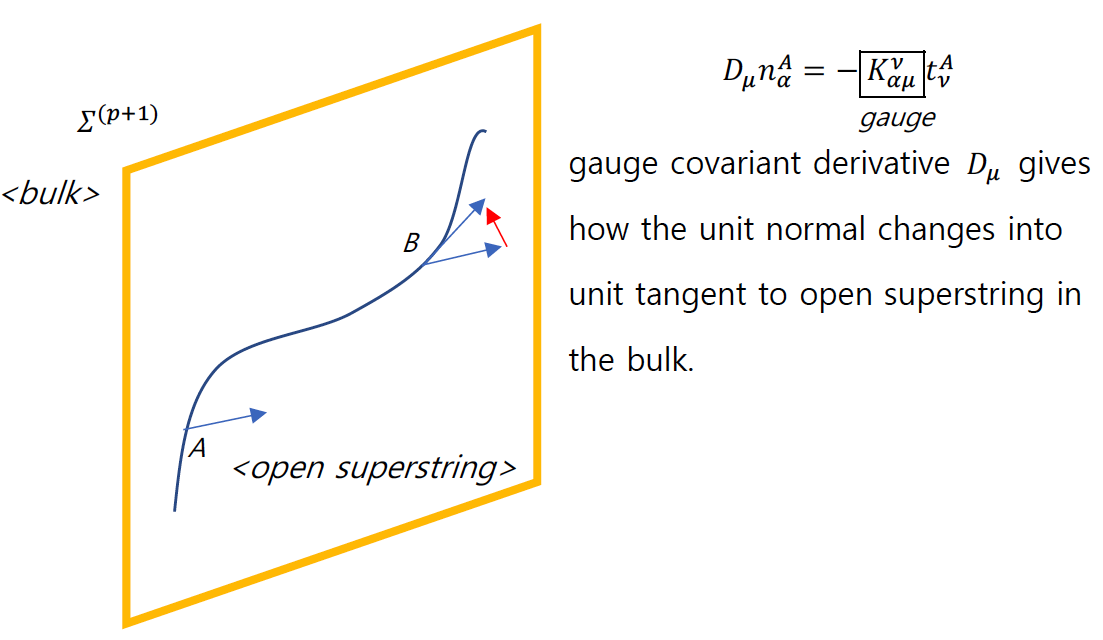}
\caption{perturbations on an open superstring produce gauge theory alone.
\newline Slightly along the open superstring unit tangent remains the same. However, unit normal may undergo some discrepancy/deviation which can be identified with unit tangent times some tensor coefficient. And this tensor coefficient involves gauge connection which, in turn, generate (generally, (non)abelian gauge theory.}
\label{fig:4}
\end{figure}

Lastly, note that the gauge covariant derivative $D_\mu$ gives how the unit normal to the open superstring changes into unit tangent in the bulk
\begin{align}
D_\mu n^A_\alpha=-K^\nu_{\alpha\mu}t^A_\nu.
\end{align}
One can now obtain the expression for $A_{\mu\alpha\beta}$
\begin{align}
&D_\mu n^A_\alpha=\partial_\mu n^A_\alpha+A^\alpha_{\mu\beta}n^A_\beta=-K^\nu_{\alpha\mu}t^A_\nu\\
&n^A_\gamma \partial_\mu n^A_\alpha+A^\alpha_{\mu\beta}(n^A_\beta n^A_\gamma)=-K^\nu_{\alpha\mu}(t^A_\nu n^A_\gamma)\\
&A^\alpha_{\mu\gamma}=-n^A_\gamma \partial_\mu n^A_\alpha=-\partial_\mu(n^A_\alpha n^A_\gamma)+n^A_\alpha\partial_\mu n^A_\gamma=n^A_\alpha\partial_\mu n^A_\gamma
\end{align}
And the field strength tensor of $A_{\mu\alpha\beta}$ is given by:
\begin{equation}
F_{\mu\nu\alpha\beta}=\nabla_\mu n^A_\alpha\nabla_\nu n^A_\beta-\nabla_\mu n^A_\beta\nabla_\nu n^A_\alpha.
\end{equation}
Now the $D_p$-brane geometry version of Gauss-Codazzi equations lead to: suppose the target spacetime geometry is that of near horizon of extremal (BPS) $D_p$-brane solution of D-dim. SUGRA, namely $M^D=AdS_{(p+2)}\times S^{D-(p+2)}$
\begin{align}
&^{(p+1)}R=0\\
&^{(p+2)}R_{\text{Ads}}=\text{const.}
\end{align}
Particularly note that although $^{(p+1)}R=0$, the three terms, $^{(p+1)}R,\ (K^\alpha_{\mu\nu}K^{\alpha\mu\nu})$,\\ $(g_{\mu\nu}K^{\alpha\mu\nu})^2$ all involve the same number of derivatives (quadratic in derivatives) and hence have equal contributions to some effective theory of $D_p$-brane.\\
 (in the above, $p=1$ as we only consider the superstring)
 
To summarize, therefore, open superstring has only vibrations/oscillation mode
whereas the spinning closed superstring has both vibration/oscillation mode and spin/winding mode.
As a result, open superstring perturbation (vibration) can only produce gauge fields whereas closed superstring perturbations \{vibration \& spin (winding)\} may well produce both gravity (=geometry) and gauge field altogether.\\

\noindent\textbf{Note}

One may exercise objection that open superstring is as fundamental as closed superstring is and hence cannot be completely ignored. As our defense, note carefully that the collision/clustering of closed superstrings would produce open superstrings instantly/repeatedly. Indeed, a hidden role of the open superstring on the D-brane upon closed superstring's scattering off the D-brane is discussed below. The discovery of this new paradigm summarized well in [Breakthrough]1, 2 in the introduction above should be recapitulated as: the gauge/gravity duality from brane/bulk duality that lasted for the past 20 years (or more). "String duality/brane physics" being replaced by; $\Rightarrow$ "string duality/spinning closed superstring".\\ 

\noindent\textbf{Paradigm change}

As the spin mode of closed superstring involves gauge theory and hence instantly replaces the quest for D-brane, after all, superstring theory \cite{1, 2, 3, 4, 5, 6, 7} being replaced by; "Closed superstring" theory and it really is true/genuine Quantum Theory Gravitation.\\

\subsection{QGD (Quantum Geometro Dynamics)}
\noindent\textbf{D-brane; its new role}

In the above, we claimed that as long as we put spin on the closed superstring, the quest for Polchinski's D-brane becomes no longer indispensable and hence literally disappears.
After all, however, if we remind that our eventual objective is to build the comprehensive theory of quantum gravitation in the context of superstring theory, the presence of Polchinski's D-brane admits a whole new perspective. In the present subsection, therefore, we would like to address this issue.

We begin with the following identifications (=);
\begin{itemize} 
\item Polchinski's D-brane = interaction vertex,
\item closed superstring ('s low-lying mode) = graviton,
\item open superstring = coupling parameter,
\item energy-momentum transfer = imprint in geometry $\&$ topology.
\end{itemize}

\begin{figure}[h!] 
\includegraphics[width=9.5cm]{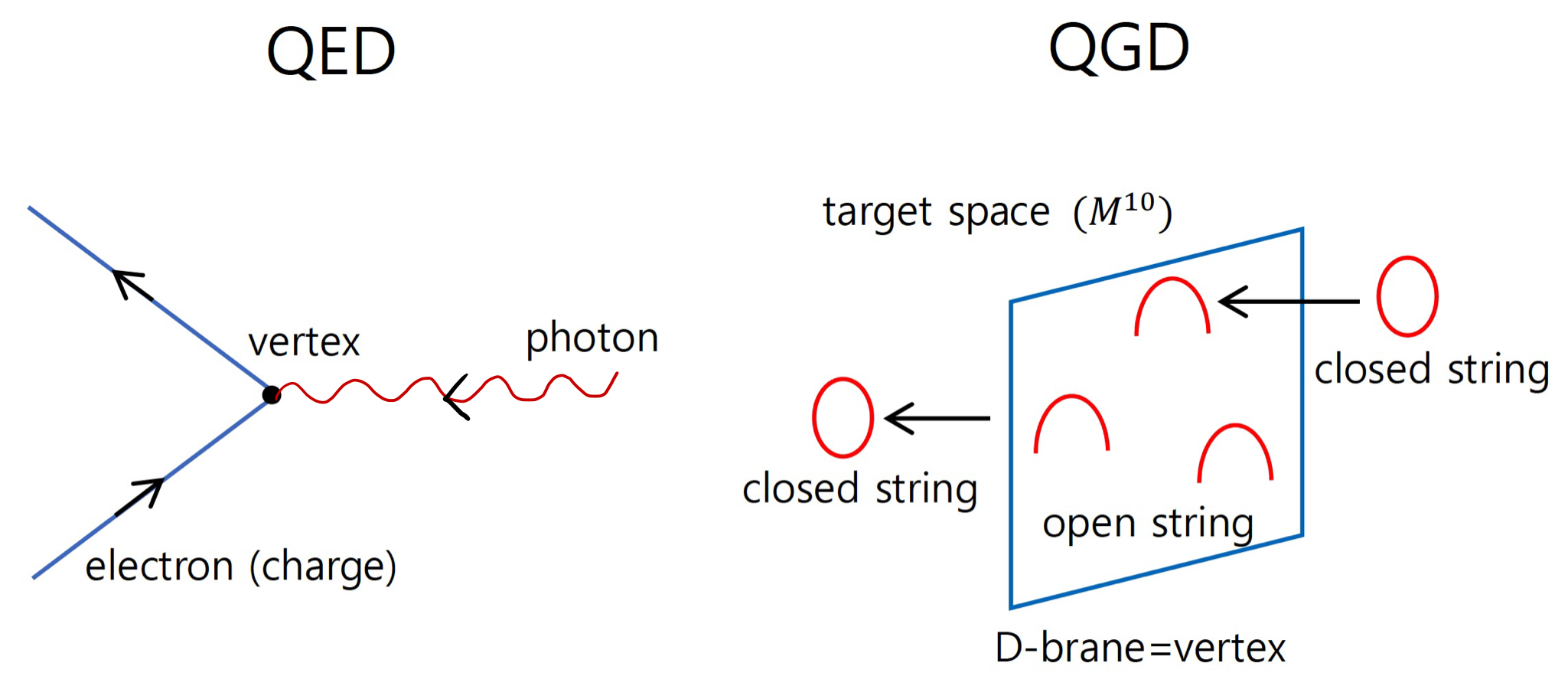}
\caption{quantum geometro dynamics (QGD); in terms of Feynman-type diagram when compared in parallel with quantum electro dynamics (QED).}
\label{fig:5}
\end{figure}
 Upon scattering off the spacetime at the vertex, i.e., D-brane, closed superstring imparts or extracts energy $\&$ momentum$\sim$radiation reaction or radiation damping by gravitational wave ($\sim$ graviton before quantisation) which, in turn, results in the change in geometry$\&$topology of the contact spacetime! 

\subsection{Summary}

\noindent{\bf{I.}}\\
Vibrations/oscillations of open superstring produce (non-abelian) gauge fields/theory.\\
$\rightarrow$ unit tangent vectors to open superstring remain the same.\\
However, unit normal vector to open superstring undergo changes.\\
$\rightarrow$ their parallel transport/total (covariant) derivatives produce (non-abelian) gauge field/theory.\\\\
{\bf{II.}}\\
Vibration/oscillations of closed superstring produce geometry/gravitation.\\
$\rightarrow$ unit normal vectors to closed superstring remain the same.\\
However, unit tangent vectors to closed superstring undergo changes.\\
$\rightarrow$ their parallel transport/total (covariant) derivatives produce geometry/gravitation.\\\\
{\bf{III.}}\\
Spin/rotation of closed superstring produces  (non-abelian) gauge fields/theory in the presence of spin (with respect to principal axis) of closed superstring\\
$\rightarrow$ unit tangent vectors to closed superstring remain the same.\\
$\rightarrow$ unit normal vectors to closed superstring undergo changes.\\
$\rightarrow$ parallel transport/total (covariant) derivatives produce (non-abelian) gauge field/theory.

To summarize/conclude, vibrations and spin of closed superstring produce geometry/gravitation and (non-abelian) gauge field/theory respectively. That is, perturbations of spinning closed superstring produce all species of particles/interactions.
Nature's one and only, exclusive building block for the entire universe is simply/just perturbations of spinning-closed superstring.
To summarize I. through III., in the presence of perturbations of superstring, or equivalently, under parallel transport $\sim$ covariant derivative of unit normal/tangent vector to the superstring.

\begin{figure}[h!] 
\includegraphics[width=5cm]{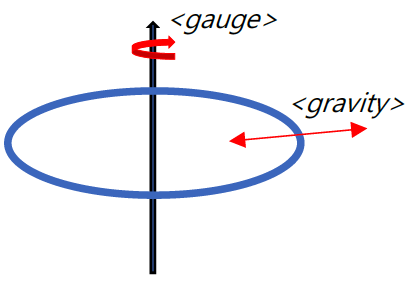}
\caption{closed superstring perturbations {vibration \& spin(winding)} produce both gravity(=geometry) and gauge theory}
\label{fig:6}
\end{figure}

\begin{itemize}
\item[1.] for an open superstring,\\
unit tangent vector remains the same,\\
unit normal vector changes and produces (non-abelian) gauge field/theory.
\item[2.] for a non-spinning-closed superstring,\\
unit normal vector remains the same,\\
unit tangent vector changes and produces (curved) geometry/gravity.
\item[3.] for a spinning-closed superstring,\\
unit normal vector changes and produces (non-abelian) gauge field/theory\\
unit tangent vector changes and produces (curved) geometry/gravity.
\end{itemize}

Therefore, putting 1, 2, 3 altogether, the spinning-closed superstring alone is nature's exclusive building block that produces both (non-abelian) gauge field/theory and (curved) geometry/gravity by its spin and vibration/oscillations respectively.\\\\

\section{MISSION 2 OF THE WORK}
\subsection{Motivations}
Ever since the advent of string paradigm that replaces the point particle description of fundamental objects, the string paradigm has undergone two major stages of development. The first stage, dubbed the first string revolution \cite{8} consists of two developments. The first development is to endow the string paradigm with the supersymmetry and the second development involves study of realistic phenomenology \cite{9} that the superstring theory \cite{10} brings.

The second stage, dubbed the second string revolution \cite{11} consists of two developments as well.
The first development is the discovery of 5 equally probable contexts of the superstring theory and the string dualities \cite{11} that bridge among these 5 species of the string theory and the second development involves another avenue which has been dubbed the “brane physics” as it has brought us a simple and reasonable way of unifying the gravitation with the rest of the three quantum gauge interactions.

As is well known, the string community conventionally employed the toroidal compactification without studying it in any serious fashion thus far. According to the result of our study given below, however, it appears to teach us that although highly complex in their nature, it is rather like the family of Calabi-Yau \cite{12, 13, 14, 15, 16} or $K3$ \cite{17} manifolds (3 fold) than the 6 torus that the nature is more likely to select. Our rationale for this claim indeed depends on the energy argument as spontaneously the nature would select the energetically favored route, that is, the Calabi-Yau or $K3$ compactification pathway. We claim that the Calabi-Yau or $K3$ manifolds have various types of flop transitions in their moduli space \cite{18} that they possess singularities which are more complex than the simple singularity of the torus which is just a single aperture. In the context of theoretical physics, it is safe to conclude that the presence of deep and steep singularities of the spacetime manifold generally implies the large negative energy states. Therefore, the minimum energy/ground state of the Calabi-Yau or $K3$ compactification case could be even lower than that of the torus compactification case. Now, we come to the major concern of our present work.

We would like to clearly distinguish between [A] the geometric T-duality which is of our particular concern and [B] Buscher's T-duality that transforms between supergravity solutions with one isometry of (Type IIA and Type IIB) \cite{11} or (Heterotic $E_8\times E_8$ and Heterotic $SO(32)$) \cite{11} superstring theories.

That is the case [A] is for T-duality on the background spacetime geometry in which the superstring theory itself is defined whereas the case [B] is for T-duality on the associated supergravity "solutions" that is the "offspring" of the "mother" superstring theory. Obviously, the relation between these two is unknown and hence remains a challenging issue of future research.

\subsection{Objectives}
To begin with, the motivation and objective of the present work can be stated as follows.
\begin{enumerate}
\item	We would like to add another, mathematical implication of the superstring T-duality \cite{11}, which is;\\
"3. trivial homotopy for mapping the closed string dynamics on one manifold onto the other T-dual manifold"\\
among 3-ways to display T-duality given in 4 Concluding Remarks below.
 
\item In terms of the “T-dual resonance” (that will be formally defined later on in this work) which is a brand new event during the 
process of compactification that had never been noticed and hence consequently employed in the entire development of superstring theory \cite{8, 11},
thus far, 
we now can make it crystal clear precisely when and how the
transmutation from stringy nature to point-particle nature and consequently when and how the transition from the superstring theory 
down to conventional point-particle physics actually take place.
\end{enumerate}

It is particularly noteworthy that this “T-dual resonance” that we discovered and report here in the present work for the first time through the entire development history of superstring theory is indeed what the
nature “spontaneously” selects but not what the human observer takes as 
the “decoupling limit \cite{8}” which is rather an arbitrary enforcement toward a
given purpose.
To summarize and conclude, therefore, the notion of “T-dual resonance” 
being discovered and reported for the first time in the present work through the entire development history of the superstring theory is indeed nature’s choice.
as it precisely is the event at which the superstring system takes the lowest ground energy state and hence gets stabilized most.
Put differently, in terms of this “T-dual resonance” that we discovered and report here in the present work for the very first time through the entire development history of the superstring theory, mother nature steps down
from the quantum gravity era represented by the current development of
the superstring theory to the familiar, well established physics of all the 4
interactions in late time universe.
Finally to conclude, this realization of the implications of T-duality particularly in terms of “T-dual resonance” that we have realized for the first time in the present work is precisely how the “spontaneous compactification” of extra spatial dimensions in superstring theory gets
realized during the quantum gravity era in the very early universe.
We now may safely declare that finally human’s science, i.e., particularly the current status of superstring theory stands for the long-seeking theory of quantum gravitation successfully coupled to the rest of already
quantized 3 interactions in nature.
And during the course of this triumph, what is ironically and even
nonsensically exciting is the realization that it is precisely the
process of compactification, the least seriously taken research issue
during the entire development history of the superstring theory
that really holds the key for the hardest issue that qualifies the superstring theory as a self-consistent and moreover, self-contained theory of
quantum gravitation that the entire theoretical physics community has been looking for over a long period of time.
At any rate, though, finally we are done.
And finally it is interesting to realize that for both the “spontaneous compactification” that we discovered here in the present work and the missing “Stringy Inflation” in the early universe that we have dealt with elsewhere which both are the long-seeking mysteries in the context of the present status of the superstring theory have been resolved by total ironies; such as compactification, the least seriously taken research issue
and the “tachyon”, the ugliest feature of any theoretical physics
respectively.

\subsection{Selection of compactification manifold}
It should be possible to define closed string T-duality \cite{11} whenever the strings can be wrapped around the non-contractible cycles/manifolds as far as the gravity \& strings is concerned. That is, when the closed/open strings are wrapped around the contractible cycles/manifolds, it would not be meaningful or even possible to define the string T-duality to begin with.
As a result, the compactification manifolds that one should consider may involve:  closed, compact \& non-contractible complex manifolds like: torus \cite{11} and Calabi-Yau/$K3$ ... but no more, that is, the string paradigm or the superstring theory itself is highly constrained on its background gravity, i.e., spacetime geometry.

As a result, for the current research issue at hand: "Implications of T-duality in superstring theory".
We take a closer look at the following selection of 3 species of spacetime geometry, that is, the compactification manifolds:
\begin{itemize}
\item[(A)] the torus compactification,
\item[(B)] the Calabi-Yau compactification, and
\item[(C)] the $K3$ compactification, respectively.
\end{itemize}

(A) For the torus compactification, torus has a singularity, that generates the minimum energy/ground state of the string system.
Indeed, in reference \cite{11} which is J. Polchinski's latest string text book, the conventional selection of this toroidal compactification has been briefly described. It is apparent that at the "T-dual resonance", $r=l_s\sim l_{pl}$, the minimum mass-squared which is roughly the energy of the closed superstring system takes place and it strongly supports the major claim of the present work of ours. To summarize, now we may "claim" that both the Calabi-Yau or $K3$ Compactification being studied in the present work in some detail by us and the conventional toroidal compactification selected by the string community thus far may admit "the minimum energy/ground state of the string system" at the T-dual resonance", $r=l_s\sim s_{pl}$ and therefore turn out to be equally favorable.
This last conclusion may sound rather discouraging as it implies no "advantage" for selecting the Calabi-Yau or $K3$ compactification over the torus compactification.
We therefore discuss this issue/fact in some more detail in the present work for the first time in the long- standing [1983 ~2018], 35 year's old  superstring physics history.

(B) For the Calabi-Yau compactification,  
as Paul S. Aspinwall,  Brian R. Greene and David R. Morrison have carried out thus far, deform/squash the geometry/topology of the Calabi-Yau manifold in order to get: "flop transition-tearing the fabric of space" then we now have a singularity, which in turn would generate the minimum energy/ground state of the string system.
\begin{figure}[h!]
\includegraphics[width=8.7cm, angle=0]{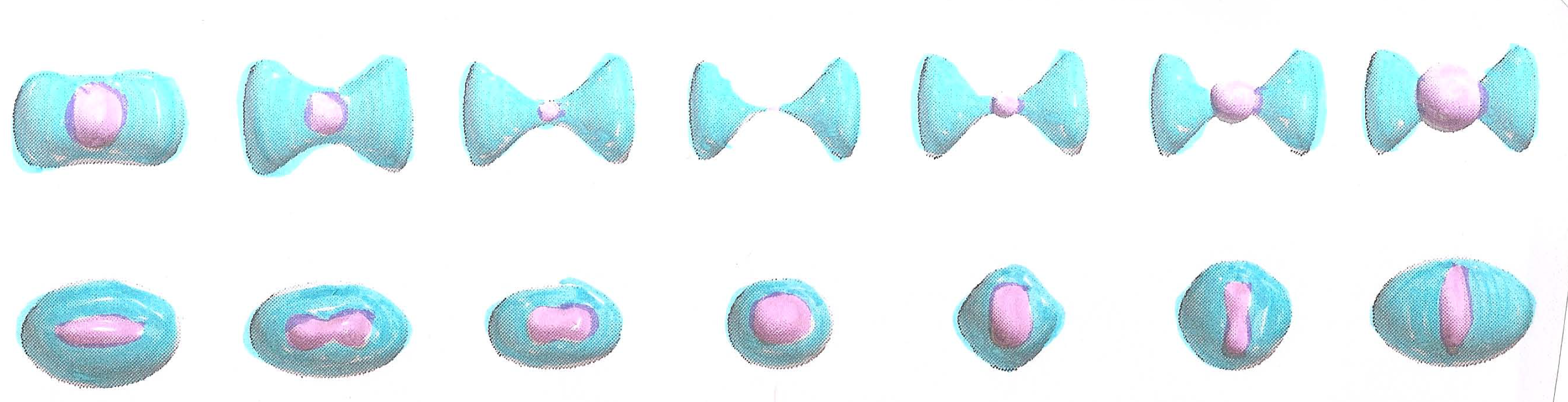}
\caption{singularities and the topology change in Calabi-Yau manifolds (courtesy B.R. Greene \cite{18}) }
\label{fig:7}
\end{figure}
Then, the above action is similar to the corresponding action to generate the "torus" in the conventional toroidal compactification selected by the string community thus far.

(C) Remind that upon defining the K$\ddot{a}$hler manifold (see Appendix \ref{appendix:a} of this article) as stated below, we now construct the $K3$ manifolds by imposing further conditions/restrictions.
By demanding the Ricci-flat K$\ddot{a}$hler manifold to have vanishing 1st Chern class or equivalently by demanding its Hermitian metric to have $SU(3)$ holonomy, one arrives at the Calabi-Yau manifolds. This additional restriction on the Ricci-flat K$\ddot{a}$hler manifolds now allows us to define the $K3$ manifolds instead. That is, starting again with the Ricci-flat K$\ddot{a}$hler manifold, if one deforms/squashes the geometry/topology in a slightly different fashion from those for the Calabi-Yau manifolds, one may instead arrive at the $K3$ manifold. This is of course well known in the algebraic geometry community in mathematics.\cite{17} As a result, roughly speaking, one now has a slightly different species of a complex manifold, called, the $K3$ spaces which have consequently a different moduli space from that of the Calabi-Yau  manifolds. It is therefore, comprehensive to consider the $K3$ compactification as well. As far as their geometry and topology are concerned, therefore, the $K3$ case would be essentially the same as those for the Calabi-Yau manifolds. To summarize, the T-duality on both the $K3$ manifolds and the Calabi-Yau would have essentially the same effect on the two manifolds.

As the major claim/ new ingredient of the present work is: the claim and its consequence that  ‘ nature’s one and only ultimate
building block is spinning-closed superstring’. As a result, the target spacetime is not  $M_{10}$ but $M_{11} = R_1\times M_{10}$,
where the  fibre manifold $R_1$ at every point of the base manifold $M_{10}$ is to accommodate  the spinning closed superstring.
This last point might need a short rationale. That is, when the closed string possesses its spin, it should be endowed with one more spacelike dimension attached to it. To be more concrete, in the absence of the spacelike dimension along the spin direction of the closed string, the presence of its spin can not be detected, to begin with. 
Then the compactification manifold consequently has to be: not $T6$  but $S^1\times T6$, not $CY3$  but $S^1\times CY3$, not $K3$  but $S^1\times K3$.

We conclude that the new compactification manifold consequently could be identified with:
‘Fibre bundles’  having the fibre manifold $S^1$ over the base manifold $M_{10}$  for all cases;
$S^1\times T6,\ S^1\times CY3,\ S^1\times K3$    respectively.

\subsection{Brief summary}
\noindent\textbf{Compactification manifolds}

As we summarize this subsection, we would like to point out that after all, it is the Calabi-Yau manifold that does it all.
\begin{align*}
M_{10}&\rightarrow C \times M_4,\\
C&=\{T6,\ CY3,\ K3\},\\
CY \text{space}&=\text{Ricci-flat \& K$\ddot{a}$hler manifold},\\
R_{\mu\nu}&=0\Rightarrow \{R=0,\ G_{\mu\nu}=0,\ R_{\mu\nu\alpha\beta}=0\}.
\end{align*}
This means that once we select the Calabi-Yau space as the compactification manifold, its scalar curvature, Einstein tensor, and Riemann curvature tensor all vanishes respectively. This point, in turn, implies the following.
As a result, Lagrangian (dynamics)=0,  Hamiltonian (energy)=0 for the compactification manifold, selected.
Gravity system is in minimum, ground state. As any (string) nature favors the lowest energy state as compactification manifolds, $CY3$ is highly favored. Besides, at the T-dual resonance, $r=\frac{l^2_s}{r} \Rightarrow r=l_s \sim l_{pl}$ the size/radius of $CY3$ is $r=l_s \sim l_{pl}$ $\Rightarrow$ closed string bounces off the $CY3$ and collapses to end up being a point particle.

\subsection{Spin on the closed superstring theory and M-theory}
As long as one puts the spin on the closed strings the closed string generates gravity from its vibrations and all the 3 gauge interactions from its spin motion. Therefore Joseph Polchinski's "D-brane" was not indispensable to begin with, as far as the unification of gravity \& gauge theories is concerned.

M-branes, meanwhile, which are simply soliton solutions in $D=11$ SUGRA which is a low-energy effective theory of still mysterious M-theory in $D=11$ still remains to be explored associated with this spinning closed strings.
Generally speaking, putting the spin onto a quantum demands (introduction, open-up) one more spacelike dimension. As a result, if one admits that it is the spinning-closed superstring which is indeed nature's one and only ultimate building block, then now the genuine critical dimension for the spinning closed string has to be elevated from $D=10$ to $D=11$ and then the M-theory which accommodates both spinning-closed superstring and the M-branes which is mother/ancestor of D-brane finally reveals/uncovers its true nature as being the full quantum theory of spinning-closed superstring which is the long-seeking full quantum theory of gravitation. It is $D=11$ spinning-closed superstring theory \& M-branes but not $D=10$ non-spinning superstring theory \& D-branes that is the true full quantum theory of gravitation. 
Why? It is $D=11$ but not $D=10$ in which the full quantum theory of gravitation which consists of spinning-closed superstring theory \& M-theory should be defined?\\

\noindent\textbf{Rationale}

Having/Seeing one more (spacelike) dimension (which could have been hidden) is indeed treating/dealing with even smaller length scale, higher energy scale. As $c\Delta t \times \Delta E \sim \Delta x \times c\Delta p=\text{constant}$, therefore at even higher energy scale, even shorter length scale, one sees/has to deal with another/hiding, hidden (spacelike) dimension.

In short, M-theory is a full quantum field theory for
\begin{itemize}
\item[(I)] spinning-closed superstring in perturbative spectrum,
\item[(II)] M-branes in non-perturbative spectrum.
\end{itemize}
True nature/identity of the M-theory in $D=11$ that has remained a (long-standing) mystery.

To summarize, M-theory in $D=11$ is a quantum field theory of spinning-closed superstring(i.e., string field theory) of which the Lagrangian (action) is that of $N=2$ super Yang-Mills theory [Seiberg-Witten theory] such that\\
(I) spinning-closed superstring is in perturbative spectrum,\\
(II) M-branes ($M_2$ \& $M_5$) are in non-perturbative spectrum.
$D=11$ M-branes, spinning closed superstring\\
$D=10$ D-branes, non-spinning closed superstring\\
As we come down from $D=11$ to $D=10$, or equivalently, at greater length scale and lower energy scale the spin of the closed superstring actually disappears (loses its meaning) and its role to unify the gravitation with the rest of the 3 quantum gauge theories is taken over by the D-branes (which are the dimensional reduction of the M-branes) as these D-brane (unlike the M-branes) possess the action of gauge-gravity duality in terms of open-closed string duality. Upon the Kaluza-Klein dimensional reduction from $D=11$ down to $D=10$.\\
In short, the one more spacelike dimension ($D=11$) allows us to discover/complete the full quantum field theory of spinning-superstring (i.e., string field theory) which has remained unknown thus far.

\subsection{Underlying Principles}
We now provide the demonstration of the major claim/conclusion of the present work. To this end, the simplest way is to refer to the expression for mass-squared of the closed superstring system given in eq.(8.3.2a) in reference \cite{11}
\begin{equation}
m^2=\frac{n^2}{R^2}+\frac{w^2R^2}{\alpha'^2}+\frac{2}{\alpha'}(N+\tilde{N}-2)
\end{equation}
 which is J. Polchinski's latest string text book. Aside from the last constant term, it is apparent that $m^2\ge2\frac{nw}{\alpha'}=2\frac{nw}{l^2_s}$ where $n=p$ and $w$ denotes the momentum number and winding number respectively and we used $\alpha'=l^2_s$ which is of order $l^2_{pl}$, namely the Planck length squared.\\
Particularly note here that the equality, that is, the minimum mass-squared takes place when $r=l_s\sim l_{pl}$ provided $w=n=p$. Here we emphasize that the minimum mass-squared which is roughly the energy of the closed superstring system takes place when: 
\begin{enumerate}
\item $r=l_s$, that is, the size of the compactification manifold (e.g., $\text{torus}, CY3, K3$) is of the string length, namely, the Planck length, as well as
\item $w=n=p$, that is, the winding mode/number becomes the same as the momentum mode/number.
\end{enumerate} 
Note that the condition 1 is so natural and hence convincing. That is, roughly, the energy of the closed string system gets minimized when the radius (i.e., the size) of the compactification manifold is precisely the string length or the Planck length.
Next, the condition 2 is not simply natural. Rather, it brings the "resolution" for the long-standing, vague conjecture or anticipation of the entire string community. That is, at the point when the energy of the closed string system gets the minimum, simultaneously $w=n=p$, which implies that the closed superstring loses its 'stringy' structure (i.e., winding) and 'transmutes' to the point particle in its internal structure. This is exactly our proof or exhibition that at the so-called, "T-dual resonance", $r=l_s\sim l_{pl}$, a generic size of the compactification manifold, the transmutation from the stringy internal structure to the conventional point-particle structure actually takes place.
To summarize and conclude, this result at which we have arrived so far is 
\begin{enumerate}
\item a fair outcome as it is precisely what we could naturally anticipate, but
\item , a gift and a relief, as it exactly coincides with entire string community's long-standing but unproven conjecture or anticipation.
\end{enumerate}
In the above, we have defined the T-duality resonance which is indeed a new ingredient that has never been referred to in superstring theory literature or the superstring community.
In short, T-duality “resonance” is the point when the string-point particle transmutation/transition takes place.

Now consider, 3-ways to display T-duality:
\begin{enumerate}
    \item $l^2_s/r\leftrightarrow r$ (exchange between the compactification manifolds with small and large radius, where $l_s$ denotes string length scale)
    \item $w=p$ (exchange between the winding number and momentum (say of the closed string))
    \item trivial homotopy for mapping the closed superstring dynamics on one manifold onto the other T-dual manifold
\end{enumerate}

\begin{figure}[h!]
\includegraphics[width=9cm]{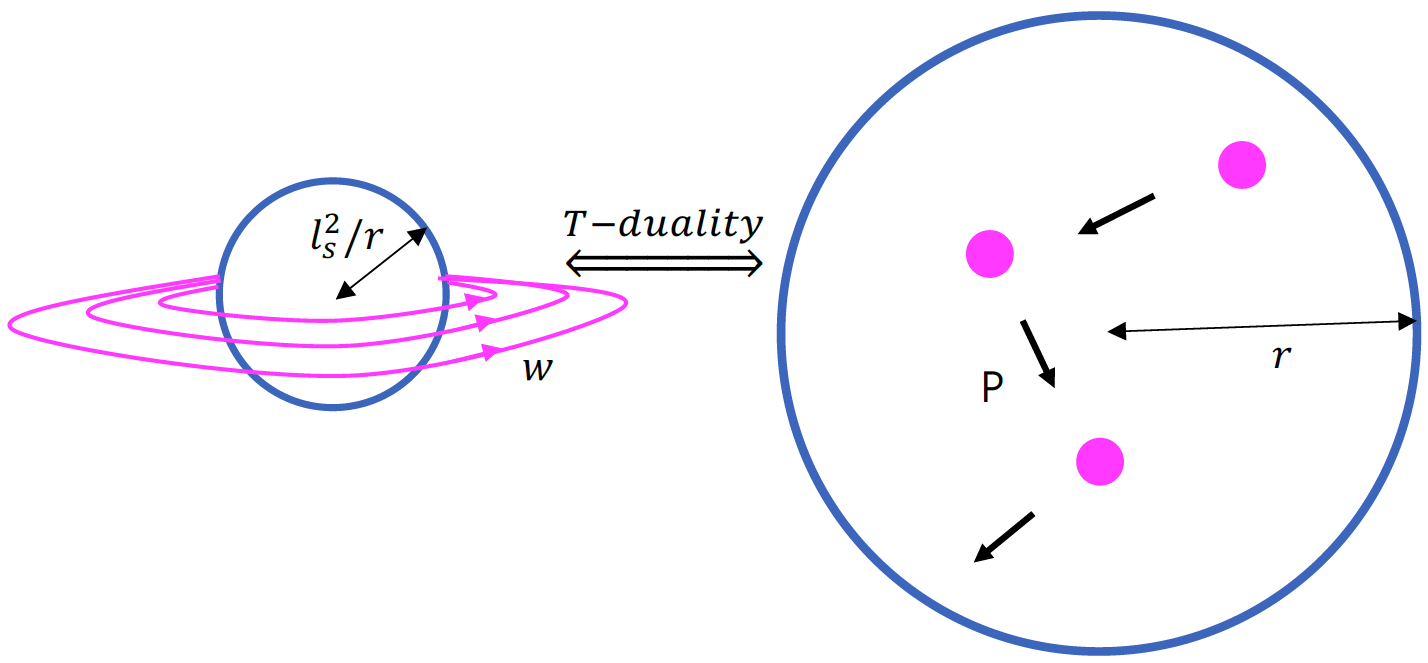}
\caption{the graphic version of the T-duality display 1, 2 above} 
\label{fig:8}
\end{figure}

\noindent We coin the case when
\begin{align*}
r=&l_s^2/r\\
r=&l_s
\end{align*}
the T-duality “resonance”.
Note that this is the minimal/shortest length scale that can be defined.\\\\
\textbf{Implication/Meaning}

To summarize, now we may "claim" that both the Calabi-Yau or $K3$ compactification studied by us in detail and the conventional toroidal compactification selected by the string community thus far may admit "the minimum energy/ground state of the string system" and therefore turn out to be equally favorable. This last conclusion, however, may sound rather discouraging as it implies no "advantage" for selecting the Calabi-Yau or $K3$ Compactification over the Torus compactification.
We therefore address this issue/fact in some more detail for the first time in the long- standing [1983 - 2018], 35 year's old  superstring physics history.

At the point when the radius of the compactification manifold (say, $CY3$, or $K3$) becomes $r=l_s$, namely, the Planck length, any winding (all windings) continuously deforms to a point (i.e., zero winding)
and hence the two, large and small compactification manifolds ($CY3$s, or $K3$) should become the same and consequently, all the winding mode collapse to a point (zero-winding).
Note particularly that at this point the closed string collapses to a point particle and hence the string’s character disappears, that is, the string-point particle transition/transmutation happens which is, interestingly, when $r=l_s$.

To summarize, the “T-dual resonance” is the point/ situation when all the compactification manifolds (say all $CY3$s) literally collapse to a point (as its radius is the smallest possible value $r=l_s$) and as a result, all the string windings or the closed strings become/collapse to a “point particle”, that is, when the closed superstring-point particle transition/transmutation takes place.

Now, starting again with the Ricci-flat K$\ddot{a}$hler manifold, if one deforms/squashes the geometry/topology in a slightly different fashion from those for the Calabi-Yau manifolds, one may instead arrive at the $K3$ manifold.
As far as their geometry and topology are concerned, therefore, the $K3$ case would be essentially the same as those for the Calabi-Yau manifolds. To summarize, the T-duality on both the $K3$ and the Calabi-Yau manifolds would have essentially the same effect on the two manifolds.

As a result, at the T-dual resonance the closed string collapses to a point particle and hence the string’s character disappears, that is the string-point particle transition/transmutation happens.
At the “T-dual resonance” when the 6-dimensional compactification manifold takes the smallest/shortest size/scale that could be defined, namely, the string scale which is nothing but the Planck scale, two remarkable events appear to
take place:
First, the string system reaches the lowest/ground energy state and hence would be stabilized on the one hand and
second, (particularly,) the closed string essentially loses all of its ‘stringy’ structure
and undergoes transmutation/transition to the particle-like internal structure that
indeed implies the actual transition of the superstring theory at the quantum gravity era toward the classical point particle physics epoch on the other hand.

Amazingly, this is what the nature’s selection of minimal size/scale superstring compactification manifold like Calabi-Yau space brings about.
Note also that this nature’s selection of minimal size/scale superstring compactification manifold like Calabi-Yau space is itself the so-called,
“spontaneous compactification” as the advent of such smallest possible compactification manifold
itself implies that the process of compactification does indeed
take place literally ‘spontaneously’.
Indeed during the development of superstring theory, the 
‘mechanism’ of “spontaneous compactification”
has remained a non-tractable mystery all the time.

Lastly we exercise a caution associated with this claim/conclusion of ours:
That is, the notion or number of spatial dimensions and their actual size/scale
do not appear to interfere or even conflict with each other as even the smallest/shortest size/scale that could be defined, namely, the string scale which is nothing but the Planck scale could accommodate as many as 6 extra/hidden
spatial dimensions.
Obviously, this notion/idea is indeed beyond our conventional thinking and hence
really is a breath-taking new realization that finally authorizes the long-standing
superstring theory as a true, legitimate theory of quantum gravitation. 

To summarize and conclude, lastly, 
the current study of ours, that is: “Implications of T-duality in superstring
theory compactification” finally authorizes the long-standing superstring theory
as indeed a true, legitimate quantum theory of gravitation that the theoretical
physics community or even the science community in general has been searching for decades of time.
And it is a relief that the current status of the superstring theory that has undergone the two revolutionary epochs is indeed at a last complete stage
which demands no more substantial elaborations.

\appendix
\section{On K$\ddot{a}$hler manifold \& Calabi-Yau spaces}
\label{appendix:a}
\paragraph{K$\ddot{a}$hler manifold}
If we have a complex manifold with Hermitian metric $g_{i\bar{j}}$ we can always construct the (1, 1)-form $\Omega=g_{i\bar{j}}dz^i\wedge d\bar{z}^j$ which is called the K$\ddot{a}$hler form. A complex manifold is called K$\ddot{a}$hler particularly if it has a closed K$\ddot{a}$hler form; $d\Omega=\frac{1}{2}(\partial+\bar{\partial})\Omega=0$.

\paragraph{K$\ddot{a}$hler potential}
The Hermitian metric of a K$\ddot{a}$hler manifold can be written in terms of a derivative of a single function, the K$\ddot{a}$hler potential $\phi(z,\bar{z})$; $g_{i\bar{j}}=\frac{\partial^2\phi}{\partial z^i\partial\bar{z}^j}$.
Besides, a K$\ddot{a}$hler manifold satisfies; $2\triangle_{\text{cl}}=\triangle_\partial=\triangle_{\bar{\partial}}$.\\

Upon defining the K$\ddot{a}$hler manifold as stated above, we now construct the Calabi-Yau manifolds by imposing further conditions/restrictions;
\begin{enumerate}
    \item K$\ddot{a}$hler manifolds, which have $U(N)$-holonomy, can be further restricted if we demand that they have vanishing 1st Chern class, in which case the $U(N)$-holonomy reduces to "$SU(N)$-holonomy",
    \item To be more specific, we require that $N=1$ supersymmetry in $D=4$ dimensions be unbroken which forces us to consider manifolds with covariantly-constant spinor $\varepsilon$; ($N=1 \ \text{SUSY})\Rightarrow(D_i\varepsilon=0)\Rightarrow(\exists$ Ricci-flat K$\ddot{a}$hler manifold with vanishing 1st Chern class $c_1=0$).
\end{enumerate}
The additional condition/restriction that should be imposed on the Ricci-flat K$\ddot{a}$hler manifold has been specified in 1, 2 above.\cite{12} Now one may wonder if the two conditions 1, 2 are really equivalent or which one should be taken? This puzzle actually has been resolved by E. Calabi and S.T. Yau first by the conjecture of E. Calabi in 1957 that subsequently has been proved by S.T. Yau in 1977. As a result, this one and the same prescription to construct the Calabi-Yau manifolds has been established as the "Calabi-Yau theorem" which  can be stated as: A K$\ddot{a}$hler manifold of vanishing 1st Chern class always admits a K$\ddot{a}$hler metric of $SU(3)$-holonomy.

\end{document}